\documentclass[showpacs,preprintnumbers,amsmath,amssymb,twocolumn,prb]{revtex4}
\usepackage{amssymb,amsmath}
\usepackage{graphics}
\usepackage{epstopdf}
\usepackage{epsfig}
\usepackage{color}
\usepackage{amsfonts}
\usepackage{bm}
\usepackage{amscd}

\begin{document}

\title{Transient quantum transport in double-dot Aharonov-Bohm interferometers }

\author{Matisse Wei-Yuan Tu}
\affiliation{Department of Physics, National Cheng Kung University,
Tainan 70101, Taiwan}

\author{ Wei-Min Zhang}
\email{wzhang@mail.ncku.edu.tw} \affiliation{Department of Physics,
National Cheng Kung University, Tainan 70101, Taiwan}

\author{JinShuang Jin}
\affiliation{Department of Physics, Hangzhou Normal University,
Hangzhou 310036, China }

\author{O. Entin-Wohlman}
\email{orawohlman@bgu.ac.il}
\altaffiliation{Also at Tel Aviv University, Tel Aviv 69978, Israel}
\affiliation{Physics Department, Ben Gurion University,  Beer Sheva
84105, Israel}

\author{A. Aharony}
\email{aaharonyaa@bgu.ac.il}
\altaffiliation{Also at Tel Aviv University, Tel Aviv 69978, Israel}
\affiliation{Physics Department, Ben Gurion University,  Beer Sheva
84105, Israel}

\begin{abstract}
Real-time nonequilibrium  quantum dynamics of electrons in double-dot 
Aharonov-Bohm (AB) interferometers is studied using an exact
solution of the master equation.  The building of the coherence between the two
electronic paths shows up via the time-dependent amplitude of the AB
oscillations in the transient transport current, and can be
enhanced by varying the applied bias on the leads, the on-site energy
difference between the dots and the asymmetry of the coupling of
the dots to the leads. The transient oscillations of the transport current do not obey
phase rigidity.  The circulating current has an anti-symmetric AB oscillation in the flux.
The non-degeneracy of the on-site energies and the finite bias cause 
the occupation in each dot to have an arbitrary flux dependence as 
the coupling asymmetry is varied.
\end{abstract}

\date{May 1, 2012}

\pacs{73.23.-b, 73.63.-b}

\keywords{Quantum decoherence, open quantum systems, quantum dots,
Aharonov-Bohm effect} \maketitle

\section{Introduction}
Coherence of electronic transport through mesoscopic junctions has
been studied intensively in nanoelectronic systems.  In particular,
the interference of electron waves has been visualized in 
Aharonov-Bohm (AB) interferometers via the AB oscillations of the conductance
of a ring placed between two leads.  Following the electron injection
from the leads into the ring, the electrons
undergo a nonequilibrium transport process before a
steady interference pattern is reached.  While the steady-state AB
interference has been largely explored in the literature, the
real-time dynamics of electronic transport in AB
interferometers has not yet been fully understood.  In this paper we
study this dynamics in double-quantum-dot AB
interferometers addressing the transient AB interference under
various tunable parameters of the system.

The study of the wavy nature of electronic transmissions has been
mainly focused on the complex amplitudes of the transmitted
electrons in the scattering approach.\cite{Buttiker92,Imry02}  The
archetype model contains a single quantum dot sitting on one of the
two arms of the AB ring. A quantum point contact (QPC) placed nearby
the quantum dot has been used to study the effect of a which-path
detection.\cite{Aleiner943740} A single-dot AB interferometer has
been realized in a closed geometry\cite{Yacoby954047} and also in an
open one.\cite{Schuster97417}  Phase rigidity in a two-terminal
geometry has been experimentally
discovered\cite{Yacoby954047,Yacoby969583} and theoretically
explained.\cite{Yeyati95R14360}  The effect of electron-electron
interactions on transport through  AB interferometers has also
been explored.\cite{Hackenbroich96110,Bruder96114} A review on the early
progress can be found in Ref.~[\onlinecite{Hackenbroich01463}].
Extracting the transmission phase from the AB oscillations is
another main issue. The way continuous phase shifts (as opposed to phase rigidity)
of the AB oscillations can be induced by breaking the unitarity of the
scattering matrix and the way such phase shifts depend on the
properties of electron losses have been investigated.
\cite{Entin02166801,Aharony02115311} Likewise, ways of extracting
both the amplitude and the phase of the intrinsic transmission
amplitude from the measured conductance without opening the
interferometer have been suggested.\cite{Aharony03156802}  The
studies of  AB interferometers with two quantum dots placed on
the two arms of the ring have been focused on different issues, such
as the flux-dependent level attraction,\cite{Kubala02245301} the
effect of intradot Coulomb
interactions,\cite{Koenig013855,Jiang04076802}  inelastic
scattering with phonons,\cite{Sigrist06036804} as well as extracting
transmission phases from the current measurements using  QPC placed next to
the one of the quantum dots.\cite{Puller10256801}

\begin{figure}[ht]
\includegraphics[width=0.7\columnwidth,angle=0]{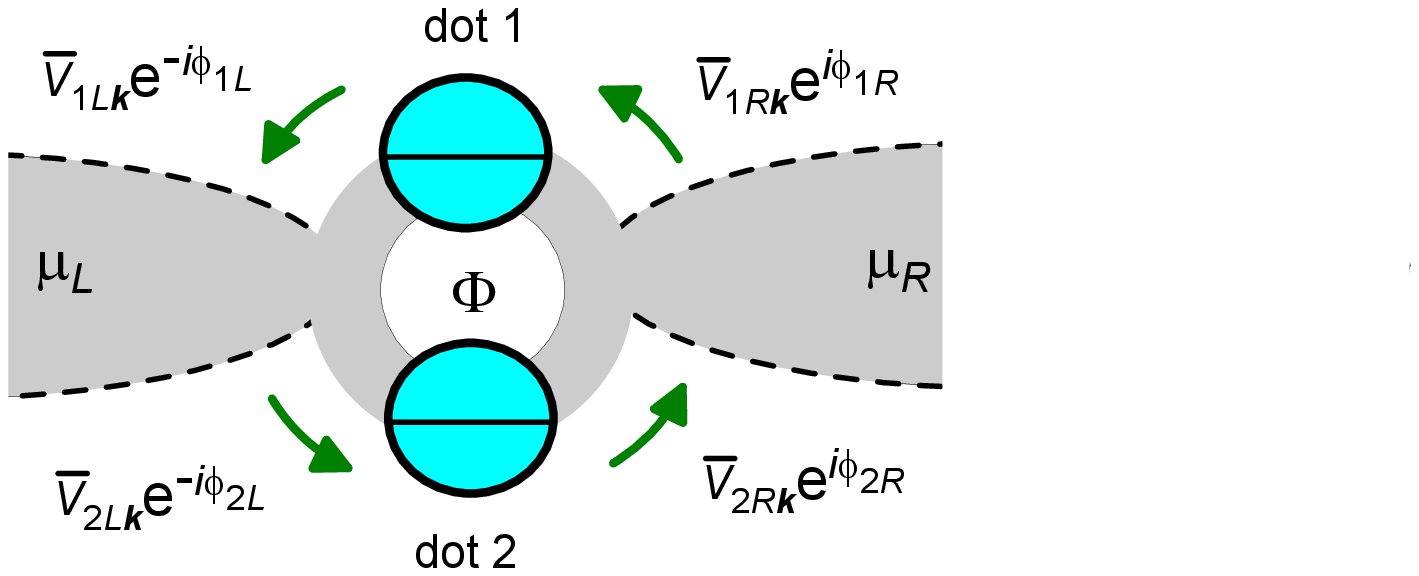}
\caption{A schematic sketch of the system: the Aharonov-Bohm
interferometer, consisting of two single-level dots, is connected to
a source and a drain set at different chemical potentials,
$\mu^{}_{L}$ and $\mu^{}_{R}$, respectively. The interferometer is
threaded by a magnetic flux $\Phi$ measured in units of the flux
quantum $\Phi_0=hc/e$.} \label{fig1}
\end{figure}

The above investigations are concentrated mainly on steady-state
properties of quantum-dot AB interferometers. In this paper we
consider the transient transport behaviors in this system. 
We consider a double-quantum-dot AB
interferometer as sketched in Fig.~\ref{fig1}, where a single active
charge state on each dot is assumed and electron-electron
interactions are ignored. In a recent work,\cite{Tu11115318}
some of us have studied the electron dynamics in this
system under the condition of identical on-site energies of the dots and
symmetric couplings to the leads. In that study, a phase localization
phenomenon has been found. In the present paper, we systematically
explore the general transient transport dynamics with non-identical on-site 
energies on the dots and asymmetric couplings to the leads. In particular,
besides the search for the dynamical flux dependences of the
transient net current, we also examine the flux dependence of the transient electronic
occupation in each dot and the transient circulating current. The electronic 
occupation on each dot can be measured in experiments and contains rich
information about the transport processes. The relatively large circulating current
at zero or small bias may provide new insights into
electron coherence during the transport.

Here is a summary of the main results we obtain. By setting the
two electronic leads (as the reservoirs) at thermal equilibrium
initially with no excess electrons on the double dot, we
monitor the time evolutions of the electronic charge occupation, the
transport net current and the circulating current. When the two
on-site energies on the dots are identical (namely the double dot
is degenerate), regardless of the coupling asymmetry to the
leads, we find that the total electronic occupation on the double dot and
the net current are always symmetric in the flux, while the occupation
difference between the two dots and the circulating current are
always anti-symmetric in it. We also find that the times needed
for the total occupation to reach its steady-state values are much longer
near zero flux, compared with the case where the flux value is away from
zero.  By breaking the degeneracy of the double dot, the net current
is allowed to break phase rigidity transiently at any bias. The flux dependence of
the total occupation number changes arbitrarily as the
the coupling asymmetry is varied at finite biases. The
non-degenerate double dot coupled asymmetrically to the leads
also drives the circulating current slightly away from
an anti-symmetric flux dependence immediately after the current is switched on,
but it then quickly becomes completely anti-symmetric in the flux.

The rest of the paper is organized as follows.  In Sec.~II we
outline the basic formalism describing the nonequilibrium 
electronic dynamics for nanoelectronic devices in general, and for the
double-dot AB interferometer in particular.  In Sec.~III we
present analytical expressions for the electronic occupations, the
transient net current and the circulating current.  In Sec.~IV we
consider the steady-state limit, reproduce known results for this system and
compare them with ours.  In Sec.~V we numerically demonstrate the
transient flux dependence of the electronic occupations and currents.
Finally, conclusions are drawn in Sec.~VI.

\section{Basic Formalism}
In this section we give a brief introduction to the
nonequilibrium quantum theory that can describe transient quantum
transport and quantum coherence in nanoelectronic
systems\cite{Tu08235311,Jin10083013} and then apply it to the 
double-quantum-dot AB interferometer considered in this paper.

The Hamiltonian of the prototypical nanoelectronic system
we consider can be written as
\begin{align} {\cal H}={\cal H}_{\rm s}+{\cal
H}_{\rm E}+{\cal H}_{\rm T}
\end{align} where
${\cal H}_{\rm s}=\sum_{ij}E_{ij}a^{\dag}_{i}a^{}_{j}$ is the
Hamiltonian of the central system with $i,j$ labeling the
electronic levels in the dots, ${\cal H}^{}_{\rm E}=\sum_{\alpha \bm{k}}
\epsilon^{}_{\alpha \bm{k}}c^{\dag}_{\alpha \bm{k}} c^{}_{\alpha
\bm{k}}$ is the lead Hamiltonian with $\alpha$ labeling the leads and
$\bm k$ denoting the states in the leads, and $ {\cal H}_{\rm
T}=\sum_{i \alpha\bm{k}}[V_{i\alpha\bm{k}} c^{\dag}_{\alpha
\bm{k}}a_{i}+{\rm H.c.}] \label{HTUN} $ describing the tunneling
between the dots and the leads. Here $a^{\dagger}_{i}$
($a^{}_{i}$) and $c^{\dag}_{\alpha \bm{k}}( c_{\alpha \bm{k}})$ are
the electron creation (annihilation) operators for electronic
levels $i$ and $\bm k$ in the dots and in lead $\alpha$,
respectively.  $E_{ii}=E_i$ is the energy of  level $i$,
$E_{ij}~(i\neq j)$ is the tunneling amplitude between the different
levels in the dots, and $V_{i\alpha k}$ is the tunneling amplitude
between the dots and the leads.  Electron-electron interactions are ignored.

Since the central system is open to the electron reservoirs
(via the leads), its nonequilibrium dynamics is naturally described by the
reduced density matrix $\rho(t)$ which is defined by tracing over the
states of the leads,
\begin{align}
\rho(t)={\rm tr}_{\rm E}\rho_{\rm tot}(t)={\rm tr}_{\rm
E}[e^{-i{\cal H}(t-t_0)}\rho_{\rm tot}(t_0)e^{i{\cal H}(t-t_0)}] ,
\label{rdm}
\end{align}
where  $\rho_{\rm tot}(t)$ is the total density matrix of
the central system plus the leads.
Electronic occupations on the discrete electronic states in the dots
can be read from $\rho(t)$.  The electronic
transport through the central system is characterized by the
currents flowing from the leads into the dots, defined by
$I_{\alpha}$=$-e\frac{d}{dt}\sum_{\bm k\in\alpha}{\rm
tr}_{tot}[c^{\dag}_{\alpha\bm k}c_{\alpha\bm k}\rho_{tot}(t)]$ for
 lead $\alpha$. This can be further decomposed into separate
contributions through each dot:
\begin{align}
I_{\alpha}=\sum_{i}I_{i\alpha}, ~~ I_{i\alpha}=ie\sum_{\bm
k\in\alpha}{\rm tr}_{\rm tot}[ V_{i\alpha\bm k}c^{\dag}_{\alpha\bm
k}a_{i}\rho_{\rm tot}(t) - {\rm H.c.}]. \label{ci}
\end{align}
In Eqs.~(\ref{rdm}) and (\ref{ci}), ${\rm tr}_{E}$ and ${\rm tr}_{\rm
tot}$ denote the traces over the states of the leads and the total
system, respectively. Throughout the paper, we use units in which
$\hbar=1$.

As usual, we assume\cite{Leg871} that the central dot system is
initially decoupled from the leads, and the leads are initially at
thermal equilibrium with the chemical potential $\mu_{\alpha}$
and inverse temperature $\beta=1/k_{B}T$ for lead $\alpha$, 
whose Fermi distribution function is given by
$f_{\alpha}(\epsilon)=1/[e^{\beta(\epsilon-\mu_{\alpha})}+1]$. Then
the exact equations governing the time evolution of the reduced
density matrix and the transient currents are
\cite{Tu08235311,Jin10083013}
\begin{subequations}
\label{rhocrnt}
\begin{align}
&\frac{d}{dt}{\rho}(t)=-i[{\cal H}_{\rm
s},\rho(t)]+\sum_{i\alpha}[{\cal L}^{+}_{i\alpha}(t)+{\cal
L}^{-}_{i\alpha}(t)]\rho(t),\label{rho}
\\&I_{i\alpha}(t)=e~{\rm tr}_{\rm s}[{\cal L}^{+}_{i\alpha}(t)]=-e~{\rm tr}_{\rm s}[{\cal
L}^{-}_{i\alpha}(t)\rho(t)], \label{crntiafa}
\end{align}
\end{subequations}
where the superoperators ${\cal L}^{\pm}_{i\alpha}(t)$ are expressed
explicitly by
\begin{align}
&{\cal L}^{+}_{i\alpha}(t)\rho(t)=-\sum_{j}\Big\{{\bm \lambda}_{\alpha
ij}(t)\big[a^{\dag}_{i}a_{j}\rho(t)+a^{\dag}_{i}\rho(t)a_{j}\big]\nonumber\\&~~~~~~~~~~~~~~~~~~+{
\bm \kappa}_{\alpha ij}(t)a^{\dag}_{i}a_{j}\rho(t)+{\rm
H.c.}\Big\},\nonumber\\& {\cal
L}^{-}_{i\alpha}(t)\rho(t)=\sum_{j}\Big\{{\bm \lambda}_{\alpha
ij}(t)\big[a_{j}\rho(t)a^{\dag}_{i}+\rho(t)a_{j}a^{\dag}_{i}\big]\nonumber\\&~~~~~~~~~~~~~~~~~~+{
\bm \kappa}_{\alpha ij}(t)a_{j}\rho(t)a^{\dag}_{i}+{\rm
H.c.}\Big\}\label{superoperators} ,
\end{align}
and ${\rm tr}_{s}$ is the trace over the states of the dots.
The first term on the right hand side of Eq.~(\ref{rho}) is
the renormalized Liouville operator of the central dot system. The
second and the third terms, expressed in terms of the superoperators, are non-unitary. 
The non-unitarity is induced by electronic dissipation and fluctuation processes
due to  the couplings of the central dot system to the electronic reservoirs.
The transient transport current is determined from the non-unitary dynamics, 
as shown by Eq.~(\ref{crntiafa}). Equations (\ref{rhocrnt}) and (\ref{superoperators})
form the basis of the nonequilibrium description of quantum
coherence and quantum transport in mesoscopic systems.

The time-dependent dissipation and fluctuation coefficients in
Eqs.~(\ref{superoperators}),
$\bm \kappa_{\alpha}(t)$ and ${\bm \lambda}_{\alpha}(t)$, are explicitly determined by the
nonequilibrium retarded and correlation Green functions of the dot
system, denoted here by $\bm u(t)$ and $\bm
v(t)$,\cite{Jin10083013} via the relations
\begin{subequations}
\label{kl}
\begin{align}
& \bm \kappa_{\alpha}(t)=\int_{t_0}^{t}d\tau {\bm
g}_{\alpha}(t,\tau){\bm
u}(\tau){\bm u}^{-1}(t) , \\
& {\bm \lambda}_{\alpha}(t)=\int_{t_0}^{t}d\tau\left\{{\bm
g}_{\alpha}(t,\tau){\bm v}(\tau)-\widetilde{\bm
g}_{\alpha}(t,\tau)\bar{\bm u}(\tau)\right\}-\bm
\kappa_{\alpha}(t){\bm v}(t) .
\end{align}
\end{subequations}
The nonequilibrium retarded and correlation Green functions of the
dot system obey the following dissipation-fluctuation
integrodifferential equations of motion
\begin{subequations}
\label{uvn}
\begin{align}
\frac{d}{dt}{\bm u}(\tau )+ i\bm E {\bm u}(\tau) +& \int_{t_0}^{\tau
} d\tau' \bm g (\tau-\tau') {\bm
u}(\tau')=0\ ,  \label{ue}\\
\frac{d}{dt}{\bm  v}(\tau)+  i\bm E \bm  v(\tau)  +&
\int_{t_0}^{\tau } d\tau' \bm g (\tau-\tau') \bm v (\tau') \notag \\
& =
 \int_{t_0}^{t }d\tau'
\widetilde{\bm g} (\tau-\tau')\bar{\bm u}(\tau')\ , \label{ve}
\end{align}
\end{subequations}
subject to the conditions $\bm
u(t_{0})=I, \bm v(t_0)=0$ with $ t_0 \leq \tau \leq t$ , and $\bar{\bm u}(\tau)={\bm
u}^\dag(t-\tau+t_{0})$ is the advanced Green function.
Here $\bm E$ is the on-site energy matrix of the dot system, ${\bm
g}=\sum_{\alpha}{\bm g}_{\alpha}$ and $\tilde{\bm
g}=\sum_{\alpha}\widetilde{\bm g}_{\alpha}$ are the self-energy
corrections due to the coupling to the leads:
\begin{subequations}
\label{gtg}
\begin{align} & {\bm
g}_{\alpha}(\tau)=\int\frac{d\omega}{2\pi}{\bm
\Gamma}_{\alpha}(\omega)e^{-i\omega\tau}, \\
& \widetilde{\bm
g}_{\alpha}(\tau)=\int\frac{d\omega}{2\pi}f_{\alpha}(\omega){\bm
\Gamma}_{\alpha}(\omega)e^{-i\omega\tau} .
\end{align}
\end{subequations}
The spectral density ${\bm \Gamma}^{}_{\alpha ij}(\omega
)=2\pi\sum_{\bm {k} \in\alpha}V_{i\alpha \bm{k}}^{*}V_{j\alpha
\bm{k}} \delta(\omega-\epsilon_{\alpha \bm k})$ summarizes  all the
non-Markovian memory effects of the electron reservoirs on the dot
system.

The correlation Green function $\bm v(t)$,
Eq.~(\ref{ve}), has a general solution in terms of the retarded Green
function $\bm u(\tau)$, 
\begin{align}
\bm v(\tau)= \int_{t_0}^{\tau}d\tau_{1}\int_{t_0}^{t}d\tau_{2} {\bm
u} (\tau-\tau_{1}+t_0) \widetilde{\bm g}(\tau_{1}-\tau_{2})
 \bar{{\bm u}}(\tau_{2}).\label{svp}
\end{align}
From the master equation, Eq.~(\ref{rho}),  it
is easy to find the single-particle reduced density matrix
in terms of $\bm u(t)$ and $\bm v(t)$:
\begin{align}
\rho^{(1)}_{ij}(t)\equiv {\rm tr}[a^\dag_j a_i \rho(t)] = u_{ii'}(t)
\rho^{(1)}_{i'j'}(t_0) u^\dag_{j'j}(t) +  v_{ij}(t) , \label{spd}
\end{align}
where $\rho^{(1)}_{ij}(t_0)$ is the initial single-particle reduced
density matrix of the dots. The currents
Eq.~(\ref{crntiafa}) can then be explicitly expressed
as\cite{Jin10083013}
\begin{align} I_{i\alpha}(t)=-2e{\rm
Re}\int_{t_{0}}^{t}d\tau\big\{{\bm g}_{\alpha}(t-\tau){\bm v}(\tau)
-\widetilde{\bm g}_{\alpha}(t-\tau)\bar{\bm u}(\tau) \nonumber \\ +
{\bm g}_{\alpha}(t-\tau){\bm u}(\tau){\bm \rho}^{(1)}(t_0) \bar{\bm
u}^\dag(t) \big\}_{ii} . \label{crntafa0}
\end{align}
This expression is consistent with the result
obtained from the Keldysh Green function technique, except that the
initial state dependence [the third term in Eq.~(\ref{spd})] is
usually ignored in most of the Green function treatments\cite{Hau98}
(see the explicit derivation given in
Ref.~[\onlinecite{Jin10083013}]). If the dot system is
initially empty, namely, $\rho^{(1)}_{ij}(t_0)=0$, the
transient electronic occupations and currents can be further simplified:
\begin{subequations}
\label{oi}
\begin{align}
&\rho^{(1)}_{ij}(t)=v_{ij}(t), \label{oe} \\
&I_{i\alpha}(t)=-2e{\rm Re} \int_{t_{0}}^{t}d\tau\big\{{\bm
g}_{\alpha}(t-\tau){\bm v}(\tau) -\widetilde{\bm
g}_{\alpha}(t-\tau)\bar{\bm u}(\tau)\big\}_{ii}\label{crntafa}.
\end{align}
\end{subequations}
Thus, solving Eq.(\ref{ue}) and using Eq.~(\ref{svp}),
we can obtain the full information of the transient quantum
transport dynamics.

To be specific, we consider in this paper a double-quantum-dot AB
interferometer schematically plotted in Fig.~\ref{fig1}, where each of the
quantum dots has a single active electronic state. Then the energy
matrix $\bm E$ in Eq.~(\ref{uvn}) becomes a $2\times2$ matrix.
We also do not consider the inter-dot tunnel coupling, namely,
$E_{12}=E_{21}=0$. The AB magnetic flux is embedded in the tunneling
amplitudes between the leads and the dots:
$V_{jL\bm{k}}=\bar{V}_{jL\bm{k}}e^{-i\phi_{jL}}$ and
$V_{jR\bm{k}}=\bar{V}_{jR\bm{k}}e^{i\phi_{jR}}$ with the relation
$\phi_{1L}-\phi_{2L}+\phi_{1R}-\phi_{2R}=\phi\equiv
2\pi\Phi/\Phi_{0}$, and $\Phi_{0}=hc/e$ is the flux quantum.  Thus the
spectral density involving explicitly the threading magnetic flux is
given by
\begin{align}
{\bm \Gamma}^{}_{\alpha  ij}(\omega )=2\pi\sum_{\bm {k}
\in\alpha}\bar{V}_{i\alpha \bm{k}}\bar{V}_{j\alpha \bm{k}}e^{\pm
i(\phi_{i\alpha}-\phi_{j\alpha})} \delta(\omega-\epsilon_{\alpha \bm
k})\ ,\label{GAM}
\end{align} where the $+(-)$ sign is for $\alpha=L(R)$.  With the above basic
formulation, we are able to explore the nonequilibrium electronic
dynamics in this nanoscale AB interferometer.

\section{Analytical solutions}
We exploit in our calculations the ubiquitously-used wide-band approximation,
in which  the spectral density is assumed to be frequency independent.
In general, the magnetic phase can be characterized by two variables,
the magnetic flux threading the ring, $\phi=\phi_{L}+\phi_{R}$,
and the difference, i.e. the gauge degree of freedom, 
$\chi=\frac{\phi_{L}-\phi_{R}}{2}$, where
$\phi_{\alpha}=\phi_{1\alpha}-\phi_{2\alpha}$.
Correspondingly, the spectral density is reduced to
$\bm \Gamma_{\alpha}=\Gamma_{\alpha}\begin{pmatrix} 1 & e^{\pm i \phi_{\alpha}}\\
e^{\mp i\phi_{\alpha}} & 1 \end{pmatrix}\ $, where the upper (lower)
sign is for $\alpha=L$($R$).  The time-dependent self-energy
correction to the retarded Green function of the electron in the
double dot is given by
\begin{align}
{\bm g}(\tau)=\delta(\tau)
\begin{pmatrix}\Gamma&e^{i\chi}\Gamma^+_\phi\\
e^{-i\chi}\Gamma^-_\phi&\Gamma\end{pmatrix}
\end{align}
with $\Gamma^\pm_\phi=[\Gamma\cos(\phi/2) \pm
i\delta\Gamma\sin(\phi/2)]$.
Here $\Gamma=\Gamma_{L}+\Gamma_{R}$ and
$\delta\Gamma=\Gamma_{L}-\Gamma_{R}$ characterize the strength and
the asymmetry of the coupling to the leads, respectively.

 Even for the most general case of a non-degenerate double
dot asymmetrically coupled to the leads, the solution of
Eq.~(\ref{ue}) can be found analytically
(taking $t_0=0$):
\begin{align}
\bm u(\tau)=u_{0}(\tau)\sigma_0-u_{p}(\tau) \hat{\vec{p}}(\phi,\chi)
\cdot \vec{\sigma} \, . \label{us}
\end{align}
Here $\vec{\sigma}=(\sigma_+, \sigma_-, \sigma_z)$ is
the vector of the three Pauli matrices, and $\sigma_0=I$
(the identity operator). We have introduced
a flux-dependent and gauge-dependent polarization vector $\vec{p}(\phi,\chi)\equiv
(p_{-}(\phi,\chi),p_{+}(\phi,\chi),p_{z}(\phi,\chi))$=$(\frac{1}{2}e^{i\chi}
\Gamma^+_\phi,\frac{1}{2}e^{-i\chi}\Gamma^-_\phi,i\delta{E})$
containing all the information on the gauge dependence, the flux
dependence and the dependence on the asymmetry of the couplings, with
$\delta{E}=E_{1}-E_{2}$ characterizing the non-degeneracy of the
double dot on-site energies. Here
$\hat{\vec{p}}(\phi,\chi)=\vec{p}(\phi,\chi)/\Gamma_\phi$
and
$\Gamma_\phi=\sqrt{\Gamma^{2}\cos^{2}(\phi/2)+\delta\Gamma^{2}\sin^{2}(\phi/2)-\delta{E}^{2}}
$ which is gauge independent.  Without loss of generality, we set
$E=\frac{E_{1}+E_{2}}{2}=0$, as an energy reference. Then the
functions $u_{0}(\tau)$ and $u_{p}(\tau)$ in Eq.~(\ref{us}) are given by
\begin{align}
& u_{0,p}(\tau)=\frac{1}{2}\left[e^{-\gamma^-_\phi\tau}
 \pm e^{-\gamma^+_\phi\tau}\right], \label{ud} 
\end{align}
with $\gamma^\pm_\phi= \frac{1}{2}(\Gamma\pm\Gamma_\phi)$, which are
also gauge independent. Substituting Eq.~(\ref{us}) into
Eq.~(\ref{svp}), we obtain the correlation Green function
\begin{align}
\bm v(t)= \int \frac{d\omega}{2\pi}{\bm u} (t,\omega)\sum_\alpha
f_\alpha(\omega)\bm \Gamma_\alpha  {\bm u}^\dag(t,\omega)\ ,
\end{align} where
${\bm u}(t,\omega) = u_{0}(t, \omega)\sigma_0-u_{p}(t,\omega)
\hat{\vec{p}}(\phi,\chi) \cdot \vec{\sigma}$ and
\begin{align}
u_{0,p}(t,\omega)=\frac{1}{2}\left[\frac{e^{(i\omega-\gamma^-_\phi)t}-1}
{i\omega-\gamma^-_\phi}
 \pm \frac{e^{(i\omega+\gamma^+_\phi)t}-1}{i\omega-\gamma^+_\phi}\right] . \label{fud}
\end{align}
The gauge degree of freedom parameterized by $\chi$ appears
explicitly in the off-diagonal matrix elements of the retarded and
the correlation Green functions ${\bm u}(t)$ and ${\bm v}(t)$. However,
the physical observables, calculated from ${\bm u}(t)$ and
${\bm v}(t)$, do not depend on $\chi$, ensuring the gauge
invariance of our calculations.

Explicitly, the electronic occupation on each dot is given by the
diagonal matrix element of $\bm v(t)$, see Eq.~(\ref{oe}).
The total occupation number $N(t)=n_{1}(t)+n_{2}(t)$, where
$n_{i}(t)=v_{ii}(t)$, can be expressed explicitly as
\begin{widetext}
\begin{align}
&N(t)=\int\frac{d\omega}{2\pi}f_{+}(\omega)\left\{\Gamma\left[
|u_{0}(t,\omega)|^{2}+(\Gamma^{2}_{\phi}+2\delta{E}^{2})|\frac{u_{p}(t,\omega)}{\Gamma_{\phi}}|^{2}
\right]-2(\Gamma^{2}_{\phi}+\delta{E}^{2}){\rm
Re}\left[\frac{u_{0}^{*}(t,\omega)u_{p}(t,\omega)}{\Gamma_{\phi}}\right]\right\}\nonumber\\&+
\int\frac{d\omega}{2\pi}f_{-}(\omega) \delta\Gamma\left\{
|u_{0}(t,\omega)|^{2}+\left(\Gamma^{2}_{\phi}+2\delta{E}^{2}-(\Gamma^{2}-\delta\Gamma^{2})
\frac{\delta{E}}{\delta\Gamma}\sin\phi\right)|\frac{u_{p}(t,\omega)}{\Gamma_{\phi}}|^{2}
-2\Gamma{\rm
Re}\left[\frac{u_{0}^{*}(t,\omega)u_{p}(t,\omega)}{\Gamma_{\phi}}\right]\right\} ,
\label{total_N}
\end{align}
\end{widetext}
and the occupation difference between the two dots,
$\delta{n}(t)=n_{1}(t)-n_{2}(t)$,  is given by
\begin{align}
\delta{n}(t)= & \int {d\omega\over2\pi}{\rm Im}
[u_{0}^{*}(t,\omega)\frac{u_{p}(t,\omega)}{\Gamma_\phi}]\Big\{\Gamma
\delta{E}f_{+} (\omega)\nonumber\\&~+\big[\delta\Gamma\delta
E-\frac{\sin\phi}{2}(\Gamma^{2}
-\delta\Gamma^{2})\big]f_{-}(\omega)\Big\}.\label{chgdif}
\end{align}
Here $f_\pm(\omega)\equiv f_L(\omega) \pm f_R(\omega)$.

On the other hand, the current passing from the left lead to the
right one through dot $i$ is given by $I_{i}=I_{Li}-I_{Ri}$.
Summing up the two currents through the two dots, we obtain the
transport net current $I=\frac{1}{2}(I_{L}-I_{R})$. Combining the
current $I_{1}$ flowing from the left to the right through the first
dot with the current $-I_{2}$ flowing from the right to the
left through the second dot gives the circulating current $I_{\rm
c}=I_{1}-I_{2}$. Explicitly, the transient net current is given by
\begin{widetext}
\begin{align}
I(t)& = \int\frac{d\omega}{2\pi}f_{+}(\omega)\Bigg\{
\delta\Gamma{\rm
Re}\Big(u_{0}(t,\omega)-\frac{\Gamma}{\Gamma_{\phi}}u_{p}(t,\omega)\Big)
 -\frac{\Gamma\delta\Gamma}{2}\Big[2|u_{0}(t,\omega)|^{2}+\Big(\Gamma^2\big[ \cos^2\frac{\phi}{2}\cos \phi +
\frac{\sin^2\phi}{2}+ \frac{\delta E}{\delta \Gamma}\sin \phi\big]
\nonumber\\& ~~~~~~~~~~~~~~~~~~~ -\delta \Gamma^2\big[
\sin^2\frac{\phi}{2}\cos\phi-\frac{\sin^{2}\phi}{2} + \frac{\delta E
}{\delta \Gamma}\sin\phi \big] +\Gamma^{2}_{\phi}+ \delta{E}^{2}
\Big)|\frac{u_{p}(t,\omega)}{\Gamma_{\phi}}|^{2}\Big] \nonumber\\&
~~~~~~~~~~~~~~~~~~~
+\Big(\Gamma^{2}\delta\Gamma(\cos^{2}\frac{\phi}{2}+1)+\delta\Gamma^{3}\sin^{2}\frac{\phi}{2}
+\frac{\Gamma^{2}-\delta\Gamma^{2}}{2}\delta{E}\sin\phi\Big){\rm
Re}\Big[\frac{u_{0}^{*}(t,\omega)u_{p}(t,\omega)}{\Gamma_{\phi}}\Big]
\Bigg\}\nonumber\\& +\int\frac{d\omega}{2\pi}f_{-}(\omega)\Bigg\{
{\rm Re}\Big(\Gamma u_{0}(t,\omega)-\frac{\Gamma^2_{\phi}+\delta
E^2}{\Gamma_{\phi}}u_{p}(t,\omega)\Big)-\frac{\delta\Gamma^{2}\left(\cos^{2}
\frac{\phi}{2}+1\right)+\Gamma^{2}\sin^{2}\frac{\phi}{2}}{2}|u_{0}(t,\omega)|^{2}
\nonumber\\& ~~~~~~~~~~~~~~~~~~~
-\frac{1}{2}\Big(\Big[\Gamma^{2}\cos^{2}
\frac{\phi}{2}-\delta\Gamma^{2}\sin^{2}\frac{\phi}{2}
-\delta{E}^{2}\Big]\delta\Gamma^{2}\cos^{2}\frac{\phi}{2}-\Big[\Gamma^{2}\cos^{2}
\frac{\phi}{2}-\delta\Gamma^{2}\sin^{2}\frac{\phi}{2}+\delta{E}^{2}\Big]
\Gamma^{2}\sin^{2}\frac{\phi}{2}\nonumber\\& ~~~~~~~~~~~~~~~~~~~
+\delta\Gamma^{2}(\Gamma^{2}_{\phi}+\Gamma^{2}\sin^{2}\phi+2\delta{E}^{2})
\Big)|\frac{u_{p}(t,\omega)}{\Gamma_{\phi}}|^{2}+2\Gamma\delta\Gamma^{2}{\rm
Re}\Big[\frac{u_{0}^{*}(t,\omega)u_{p}(t,\omega)}{\Gamma_{\phi}}\Big]
 \Bigg\}, \label{netcrnt}
\end{align}
\end{widetext}
where the dependencies on $\chi$ in the Green functions ${\bm u}$
and ${\bm v}$ are exactly canceled by those of the self-energy
corrections $\tilde{\bm g}$ and ${\bm g}$ [see
Eq.~(\ref{crntafa})], leaving the current gauge independent.
The transient circulating current is given by
\begin{align}
I_{c}(t)=& \int  {d\omega\over2\pi}f_{+}(\omega){\rm
Im}\Bigg\{  -\frac{u_{p}(t,\omega)}{\Gamma_\phi}(\Gamma^2-\delta\Gamma^2)\sin\phi
\nonumber\\& + \delta\Gamma\delta{E}\left[2\frac{u_{p}(t,\omega)}{\Gamma_\phi}-\Gamma
u_{0}^{*}(t,\omega)\frac{u_{p}(t,\omega)}{\Gamma_\phi}\right]
\Bigg\}\notag\\
+\int & {d\omega\over2\pi}f_{-}(\omega){\rm
Im}\Bigg\{  u_{0}^{*}(t,\omega)\frac{u_{p}(t,\omega)}{\Gamma_\phi}
\frac{\delta\Gamma}{2}(\Gamma^{2}-\delta\Gamma^{2})\sin\phi
\nonumber\\& + \delta{E}\big[2\Gamma\frac{u_{p}(t,\omega)}{\Gamma_\phi}-\delta
\Gamma^{2}u_{0}^{*}(t,\omega)\frac{u_{p}(t,\omega)}{\Gamma_\phi}\big]\Bigg\}.
\label{circrnt}
\end{align}  These dynamical quantities
depend on the amount of the non-degeneracy $\delta{E}$,
the coupling asymmetry $\delta\Gamma$, the magnetic flux
$\phi$ and also the bias voltage applied on the leads through the
particle distributions in the two electronic reservoirs.  The time
scales for the transient behaviors of these physical observables are
determined by the factors $1/\gamma^\pm_\phi=2/(\Gamma\pm \Gamma_\phi)$
in Eq.~(\ref{ud}), in which the flux as well as the coupling asymmetry
and the non-degeneracy play their important roles.

The symmetric and degenerate double-dot interferometer has been
widely studied in the literature. This corresponds to $\delta E= \delta \Gamma
=0$. Thus $\Gamma_\phi=\Gamma |\cos \frac{\phi}{2}|$, and the
above results can be significantly simplified.  Explicitly, the total occupation
number in the double dot is reduced to
\begin{align}
N (t) = \Gamma \int & \frac{d\omega}{2\pi}
f_{+}(\omega)\Big\{|u_{0}(t,\omega)|^{2} +|u_{p}(t,\omega)|^{2}
\notag
\\ &-2|\cos(\phi/2)| {\rm
Re}[u_{0}^{*}(t,\omega)u_{p}(t,\omega)] \Big\} ,
\end{align} and the occupation difference between the two dots becomes
\begin{align}
\delta{n}(t)= &\Gamma\sin(\phi/2)\frac{\cos(\phi/2)}{|\cos(\phi/2)|}\times\nonumber\\
&\int {d\omega\over2\pi} f_{-}(\omega) {\rm Im}
[u_{0}^{*}(t,\omega)u_{p}(t,\omega)].
\end{align}  The transient net current is simplified to be
\begin{align}
I(t)= \int \frac{d\omega}{2\pi} &
f_{-}(\omega)\Big\{-\Gamma^{2}\frac{|u_{0}(t,\omega)|^{2} -
|u_{p}(t,\omega)|^{2}}{2}\sin^{2}\frac{\phi}{2} \nonumber\\&+\Gamma
{\rm Re}[u_{0}(t,\omega) -|\cos(\phi/2)| u_{p}(t,\omega)]\Big\}
\label{netcrnts}
\end{align} and the circular current is given by
\begin{align}
I_{c}(t)=  -\Gamma \sin(\phi/2)\frac{\cos(\phi/2)}{|\cos(\phi/2)|}
\int {d\omega\over2\pi}f_{+}(\omega) {\rm Im}[u_{p}(t,\omega)] .
\label{circrnts}
\end{align}

In general, the transient flux dependence of the physical
quantities, Eqs.~(\ref{total_N})-(\ref{circrnt}), for non-degenerate
double dot with asymmetric couplings to the leads, are neither symmetric
nor anti-symmetric in the flux.  The complicated flux dependencies
are mainly determined by the energy splitting $\delta{E}$. When the
two quantum dots are set at degeneracy, $\delta{E}=0$, regardless of
the coupling asymmetry and the finite applied bias, both the total
occupation number and the net current become symmetric in the flux,
namely, $N(\phi,t)=N(-\phi,t)$ and $I(\phi,t)=I(-\phi,t)$.  In
contrast, the occupation difference and the circulating current
become anti-symmetric in the flux:
$\delta{n}(\phi,t)=-\delta{n}(-\phi,t)$ and
$I_{c}(\phi,t)=-I_{c}(-\phi,t)$. However, when the degeneracy is
lifted, both the symmetric and anti-symmetric flux dependencies are
transiently present in all these physical observables. 
These complicated flux dependencies can be simplified by
setting an applied bias, $\mu_{L}=eV/2=-\mu_{R}$.  Under such a bias
configuration, the terms involving $f_{-}(\omega)$ in
Eqs.~(\ref{chgdif}) and (\ref{circrnt}) vanish since both ${\rm Im}
[u_{0}^{*}(t,\omega)u_{p}(t,\omega)]$ and ${\rm Im}
[u_{p}(t,\omega)]$ are odd in $\omega$ while $f_{-}(\omega)$ is even
in $\omega$.  The difference in the occupations of the two dots then
becomes symmetric in the flux,
$\delta{n}(\phi,t)=\delta{n}(-\phi,t)$, and is proportional to
$\delta{E}$. On the other hand, the circulating current generally
contains two contributions. One is proportional to
$(\Gamma^{2}-\delta\Gamma^{2})\sin(\phi)$ and is anti-symmetric in
the flux.  The other contribution is proportional to
$\delta\Gamma\delta{E}$, and is symmetric in the flux. However, we
find that the second contribution decays to zero within a time scale
of a few $1/\Gamma$. After that time, the circulating current
becomes anti-symmetric in the flux, proportional to $\sin(\phi)$. 
Applying the aforementioned
bias configuration does not affect the existence of both the
symmetric and the anti-symmetric flux dependent components of the
total occupation and the transient net current. Only in the special
case of zero bias, the total occupation number becomes symmetric in
the flux.  The transient net current always has a non-vanishing
anti-symmetric flux dependence for arbitrary values of the bias when
$\delta{E}\ne0$.

More interestingly, during the transient transport processes,
$I(\phi,t)\ne I(-\phi,t)$ for the non-degenerate case. In other
words, it transiently breaks the well-known phase rigidity at
arbitrary biases. This is easily understood because during the
nonequilibrium transient processes there is no time-reversal
symmetry. The time-reversal symmetry is the prerequisite for phase
rigidity of the linear conductance of a two-terminal
device.\cite{Yacoby954047, Yacoby969583, Yeyati95R14360} Only
at steady state, as we show in the next section, can
Eq.~(\ref{netcrnt}) reproduce this phase rigidity,
independent of the value of $\delta{E}$. The on-site energy
splitting thus plays a crucial role for the time-reversal symmetry
breaking with respect to the flux dependency during the
transient dynamics.  Note also that $I(\phi,t)\ne I(-\phi,t)$ is
a purely transient effect.  In a steady state, the phase rigidity 
is preserved.

\section{Coherence and phase rigidity at steady state}

Before studying the real-time dynamics of electronic transport
in this double-dot AB interferometer, we deduce the steady-state
results from the general transient solutions given in Sec.~III,
and compare them with the previous steady-state solutions
obtained for a degenerate double dot coupled symmetrically
to the leads.

\subsection{The general steady-state solution}
Taking the steady-state limit of Eq.~(\ref{ud}), we have
$\lim_{t\rightarrow\infty}u_{0,p}(t,\omega)=\frac{1}{2}(u_{-}(\omega)
\pm u_{+}(\omega))$ where
$u_{\pm}(\omega)=\frac{1}{\gamma^{\pm}_\phi-i\omega}$.  Substituting
these solutions into Eqs.~(\ref{total_N})-(\ref{circrnt}), we obtain the
electronic occupation and transport currents at steady state.

The total electronic occupation at steady state is reduced to
\begin{align}
N(\phi ) &=\Gamma\int_{-\infty }^{\infty }{\frac{d\omega }{2\pi
}}f_{+}(\omega)\frac{(\omega^{2}+\gamma^{+}_\phi \gamma^{-}_\phi)}
{[\omega^{2}+(\gamma^{+}_\phi)^{2}][\omega^{2}+(\gamma^{-}_\phi)^{2}]}\nonumber\\&+
\int_{-\infty }^{\infty }{\frac{d\omega }{2\pi
}}f_{-}(\omega)\Big\{\frac{\delta\Gamma(\omega^{2}-\gamma^{+}_\phi\gamma^{-}_\phi)}
{[\omega^{2}+(\gamma^{+}_\phi)^{2}][\omega^{2}+(\gamma^{-}_\phi)^{2}]}\nonumber\\&
~~~~~~~~~~~~~~~ +\delta {E}\frac{\frac{%
\delta \Gamma \delta E}{2}-\frac{\Gamma ^{2}-\delta \Gamma
^{2}}{4}\sin (\phi )}{[ \omega ^{2}+(\gamma^{+}_\phi)^{2}][ \omega
^{2}+(\gamma^{-}_\phi)^{2}]}\Big\}. \label{totocinf}
\end{align}  At zero bias, only the first term survives.
At a finite bias, the difference of the particle
distributions between the two electronic reservoirs can give
an additional contribution to the total occupation
when the asymmetric coupling or the non-degeneracy are present.
Besides, it shows that the flux dependence of the total
occupation has an anti-symmetric flux dependence only when a
finite bias is applied for a non-degenerate
double dot.
The steady-state solution of the occupation difference is
\begin{align}
&\delta n(\phi )=\nonumber\\&\int_{-\infty }^{\infty }{\frac{d\omega
}{2\pi }}\frac{ \omega \left[ \Gamma \delta Ef_{+}(\omega )+\left(
\delta \Gamma \delta E-\frac{(\Gamma ^{2}-\delta \Gamma
^{2})}{2}\sin (\phi )\right) f_{-}(\omega )\right] }{(\omega
^{2}+(\gamma^{+}_\phi)^{2})(\omega
^{2}+(\gamma^{-}_\phi)^{2})}\text{.}\label{chgdif_stdy}
\end{align}  Equation (\ref{chgdif_stdy}) shows further
that one must have either an on-site energy splitting or a nonzero
flux under a finite bias to generate an occupation difference between the 
 two dots.

The general expression of the steady-state net current is 
\begin{align}
I(\phi)=\int\frac{d\omega}{2\pi}[f_{L}(\omega)-f_{R}(\omega)]
\mathcal{T}(\omega,\phi), \label{sscurrent}
\end{align}
where the transmission coefficient is 
\begin{align}
\mathcal{T}(\omega,\phi)=\frac{(\Gamma^{2}-\delta\Gamma^{2})[\omega^{2}\cos^{2}\frac{\phi}{2}
+(\frac{\delta{E}}{2}\sin\frac{\phi}{2})^2]}
{[\omega^{2}+(\gamma^{+}_\phi)^{2}][\omega^{2}+(\gamma^{-}_\phi)^{2}]}
. \label{ctm}
\end{align}
By taking $\delta\Gamma=0$ and $\delta{E}=0$, Eq.~(\ref{ctm})
reproduces the results investigated in Refs.~[\onlinecite{Kubala02245301,Koenig013855}].
Equation (\ref{sscurrent}) shows that for this two-terminal device the
steady-state net current (as well as its derivative with respect to the bias, i.e. the
differential conductance) obey phase rigidity $I(\phi)=I(-\phi)$.  
The AB flux profile of the
steady-state net current exhibits a phase shift of only 0 or $\pi$ with
respect to all possible variations of the system parameters, including
the non-degeneracy of the double dot and the asymmetry of the couplings to
the leads.  Correspondingly, the steady-state circulating current is given by
\begin{align}
&I_{c}(\phi)=(\Gamma^{2}-\delta\Gamma^{2})\times\nonumber\\&\int_{-\infty }^{\infty }{\frac{%
d\omega }{2\pi }}\frac{\omega \left[ -\frac{\Gamma }{2}\sin (\phi
)f_{+}(\omega )+(\delta E+\frac{\delta \Gamma }{2}\sin (\phi ))f_{-}(\omega )%
\right] }{[ \omega ^{2}+(\gamma^{+}_\phi)^{2}][ \omega
^{2}+(\gamma^{-}_\phi)^{2}]}. \label{stdy_Ic}
\end{align}
Both the net current and the circulating current, see
Eqs.~(\ref{sscurrent}) and (\ref{stdy_Ic}), are proportional to
$\Gamma^{2}-\delta\Gamma^{2}$.  Their AB oscillation
amplitudes decrease upon increasing the coupling asymmetry.

\subsection{Small and large bias limits at zero temperature}
The steady-state occupation numbers and currents, Eqs.~(\ref{totocinf})-(\ref{stdy_Ic}),
are expressed in terms of integrals over the frequency.  These
integrals can be explicitly carried out at zero temperature with the
bias configuration $\mu_{L}=eV/2=-\mu_{R}$.

At zero temperature, the total
electronic occupation is found to be
\begin{subequations}
\label{total_N_zerotemp}
\begin{align}
&N(\phi)=\nonumber\\& 1+\Big( \delta \Gamma -\delta {E}\frac{
\frac{\delta \Gamma \delta E}{2}-\frac{\Gamma ^{2}-\delta \Gamma
^{2}}{4} \sin \phi }{\Gamma \gamma^{+}_\phi }\Big) \frac{\tan^{-1}
\Big[ \frac{eV/2}{\gamma^{+}_\phi}\Big]}{\pi \Gamma_\phi }
\nonumber\\ &-\Big( \delta \Gamma -\delta {E}\frac{\frac{ \delta
\Gamma \delta E}{2}-\frac{\Gamma ^{2}-\delta \Gamma ^{2}}{4}\sin
\phi }{\Gamma \gamma^{-}_\phi}\Big)\frac{\tan^{-1} \Big[
\frac{eV/2}{ \gamma^{-}_\phi}\Big]}{\pi \Gamma_\phi } \text{.}
\end{align}
Assuming a small or a large bias, Eq.~(\ref{total_N_zerotemp}) can be further
simplified,
\begin{align}
&N(\phi )\rightarrow  1 + \nonumber \\
&\left\{\begin{array}{ll}
\frac{eV}{2 \pi \gamma^{+}_\phi \gamma^{-}_\phi}\Big[ \delta
{E}\frac{2\delta \Gamma \delta E-(\Gamma ^{2}-\delta \Gamma
^{2})\sin \phi }{4\gamma^{+}_\phi\gamma^{-}_\phi}-\delta \Gamma \Big]
& \mbox{if $eV \ll \Gamma $} \\ ~~ & ~~ \\
\frac{\delta{E}}{2\Gamma}\frac{2\delta{E}\delta\Gamma-
(\Gamma^{2}-\delta\Gamma^{2})\sin\phi}{(\Gamma^{2}-\delta\Gamma^{2})
\sin^{2}(\phi/2)+\delta{E}^{2}} & \mbox{if $eV \gg \Gamma $} \end{array} \right.
\end{align}
\end{subequations}
Note that since Coulomb interactions have been ignored, the screening
effect is altogether discarded.  By
setting $\delta{E}=0$, the total occupation at large bias
becomes independent of the flux. In contrast, at small bias
the total occupation is flux dependent when the
coupling to the leads becomes asymmetric.

The occupation difference between the two dots at zero temperature reads
\begin{subequations}
\label{chgdif_zerotemp}
\begin{align}
&\delta{n}(\phi)= \nonumber \\
&~~~\left\{\begin{array}{ll} \frac{\delta E}{ 2\pi \Gamma_\phi} \ln
\frac{(eV/2)^{2}+(\gamma^{-}_\phi)^2}{
(eV/2)^{2}+(\gamma^{+}_\phi)^2} & \mbox{if $\Gamma_\phi$ is~real}\\
\frac{\delta E}{ \pi |\Gamma_\phi|}\Big[\tan^{-1}
\frac{(eV)^{2}+\Gamma^2 -|\Gamma_\phi|^2}{ 2\Gamma|\Gamma_\phi|^2}
-\frac{\pi}{2}\Big]& \mbox{otherwise}
\end{array} \right. .
\end{align}
As expected, the  occupation difference is proportional to $\delta E$.
 The small and large bias limits are
\begin{equation}
\delta{n}(\phi) \rightarrow  \left\{ \begin{array}{ll} \frac{\delta
E}{\pi\Gamma_\phi}\ln  \frac{\gamma^{-}_\phi}{
\gamma^{+}_\phi}  & \mbox{if $eV \ll \Gamma $}\\
0 & \mbox{if $eV \gg \Gamma $} \end{array} \right.
\end{equation}
when $\Gamma_\phi$ is real, and otherwise
\begin{equation}
\delta{n}(\phi) \rightarrow  \left\{ \begin{array}{ll} \frac{\delta
E}{\pi|\Gamma_\phi|}\Big[\tan^{-1}
 \frac{\Gamma^2 -|\Gamma_\phi|^2}{
2\Gamma|\Gamma_\phi|^2} -\frac{\pi}{2}\Big] & \mbox{if $eV \ll \Gamma $}\\
0 & \mbox{if $eV \gg \Gamma $} \end{array} \right. .
\end{equation}
\end{subequations}
Therefore, when a small bias is applied, the on-site energy splitting
effectively causes a difference in the occupations.  However, when we 
apply a bias much larger than the energy splitting, the energy
splitting becomes ineffective in rendering the occupation difference
between the two dots. The bias setting with respect to the energy
splitting is thus essential for the control of the occupation difference 
between the two dots.

Having examined the occupations in these limits, we now turn to the
currents. The steady-state net current at zero temperature is found to be
\begin{subequations}
\label{sscurrent_zerotemp}
\begin{align}
I(\phi ) &=  \frac{(\Gamma ^{2}-\delta \Gamma ^{2})}{\pi \Gamma \Gamma_\phi}\nonumber\\
&\times\Big\{\big( \gamma^{+}_\phi\cos ^{2}\frac{\phi}{2}-\frac{\delta
E^{2}\sin ^{2}\frac{\phi}{2}}{4\gamma^{+}_\phi}\big) \tan^{-1} \big[
\frac{eV/2}{\gamma^{+}_\phi}\big]  \nonumber\\
&~~~~~-\big( \gamma^{-}_\phi\cos ^{2}\frac{\phi}{2}-\frac{\delta E^{2}\sin
^{2}\frac{\phi}{2}}{4\gamma^{-}_\phi}\big) \tan^{-1} \big[
\frac{eV/2}{\gamma^{-}_\phi}\big] \Big\} .
\end{align} 
For small or large biases, it is further reduced
to
\begin{align}
&I(\phi )\rightarrow  (\Gamma^{2}-\delta \Gamma ^{2})
\nonumber \\ &~~\times \left\{\begin{array}{ll}\frac{eV}{8\pi
(\gamma^{+}_\phi\gamma^{-}_\phi)^2}
\delta E^{2}\sin ^{2}\frac{\phi}{2} & \mbox{if $eV \ll \Gamma $} \\
\frac{1}{2\Gamma}\big[
\cos^{2}\frac{\phi}{2}+\frac{\delta{E}^{2}\sin^{2}\frac{\phi}{2}}
{(\Gamma^{2}-\delta\Gamma^{2})\sin^{2}\frac{\phi}{2}+\delta{E}^{2}}
\big] & \mbox{if $eV \ll \Gamma $} \end{array} \right. .
\end{align}
\end{subequations}
In the small bias limit, the amplitude of the AB oscillation in the
net current increases with the on-site energy splitting. However,
under a large bias, it shows two competing oscillations,
$\cos^{2}(\phi/2)$ and $\sin^{2}(\phi/2)$, which results in a
sub-oscillatory pattern over the main oscillation of
$\cos^{2}(\phi/2)$, proportional to $\delta E$.

The explicit expression for the steady-state circulating current at zero
temperature is
\begin{subequations}
\label{circrnt_zerotemp}
\begin{align}
&I_{c}(\phi) =-\frac{\Gamma^{2}-\delta\Gamma^{2}}{ 2\pi
}\sin\phi \nonumber \\
& ~~\times \left\{ \begin{array}{ll} \frac{1}{2\Gamma_\phi} \ln
\frac{(eV/2)^{2}+(\gamma^{-}_\phi)^2}{
(eV/2)^{2}+(\gamma^{+}_\phi)^2}  & \mbox{if $\Gamma_\phi$ is~real}\\
\frac{1}{|\Gamma_\phi|}\Big[ \tan^{-1} \frac{(eV)^{2}+\Gamma^2
-|\Gamma_\phi|^2}{ 2\Gamma|\Gamma_\phi|^2} -\frac{\pi}{2}\Big] &
\mbox{otherwise} \end{array} \right.  , \label{cczt}
\end{align} 
and for small or large biases it reads
\begin{align}
&I_{c}(\phi) \rightarrow \left\{\begin{array}{ll}
 -\frac{(\Gamma
^{2}-\delta \Gamma ^{2})}{2\pi \Gamma_\phi}\sin \phi \ln
\frac{\gamma^{-}_\phi}{\gamma^{+}_\phi}
& \mbox{if $eV \ll \Gamma $} \\
0 & \mbox{if $eV \gg \Gamma $} \end{array} \right. \label{ccsb}
\end{align}
when $\Gamma_\phi$ is real, and otherwise
\begin{align}
&I_{c}(\phi) \rightarrow  \nonumber \\
&~~~ \left\{\begin{array}{ll}
 -\frac{(\Gamma
^{2}-\delta \Gamma ^{2})}{2\pi |\Gamma_\phi|}\sin \phi\Big[
\tan^{-1} \frac{\Gamma^2 -|\Gamma_\phi|^2}{ 2\Gamma|\Gamma_\phi|^2}
-\frac{\pi}{2}\Big]
& \mbox{if $eV \ll \Gamma $} \\
0 & \mbox{if $eV \gg \Gamma $} \end{array} \right.  .\label{ccsb1}
\end{align}
\end{subequations}
From the above results we find that the circulating
current becomes significantly large when the bias is
sufficiently small. In the opposite limit, the large bias drives the electron
to flow in one direction and the circulating motion is then strongly
suppressed.

It is worth noting that in the case
of the degenerate double dot at zero flux,  the operator $
A^{\dag}_{-}A_{-}$, where
$A_{-}=\frac{1}{\sqrt{2}}(e^{-i\chi/2}a_{1}-e^{i\chi/2}a_{2})$ and $\chi$
is the gauge degree of freesom,
gives a constant of motion:\cite{footnote1}
$[A^{\dag}_{-}A_{-},\mathcal{H}]=0$. When one turns on a finite
flux, this symmetry is broken, and the electronic occupation is
changed significantly from the value at zero flux.  Indeed, if one sets $\delta{E}=0$ and
$\phi=0$ in Eq.~(\ref{us}) and takes the
steady-state limit, one obtains
$N(\phi=0)=1/2+\frac{\delta \Gamma }{\pi \Gamma }\tan^{-1} \left[ \frac{eV}{%
2\Gamma }\right]$ at zero temperature.  However, taking the zero
flux limit in Eq.~(\ref{total_N_zerotemp}) at degeneracy,
one finds $N(\phi\rightarrow0)=1-\frac{\delta \Gamma }{2\Gamma }+\frac{%
\delta \Gamma }{\pi \Gamma }\tan^{-1} \left[ \frac{eV}{2\Gamma
}\right]$.  This indicates that $N(\phi=0)\ne N(\phi\rightarrow0)$, namely
the total occupation, changes abruptly across the zero flux
point. On the other hand, Eq.~(\ref{chgdif}) shows that
at degeneracy
$\delta{n}(\phi=0,t)=\lim_{\phi\rightarrow0}\delta{n}(\phi,t)=0$,
namely the occupation difference $\delta{n}$ is continuous across
the zero flux point. By setting first $\delta{E}=0$ and
$\phi=0$ in Eq.~(\ref{us}) and then taking the steady-state limit,
compared with the limit $\phi\rightarrow0$
after the steady-state limit is taken, we find that both $I$ and $I_{c}$ are
continuous as the zero flux point is crossed. Thus the abrupt change
upon crossing zero flux occurs only in the total electronic occupation
due to the existence of an occupation constant of motion at
$\phi=0$.

The results presented in this section give the general AB
flux dependence of the electronic occupation and electronic transport in
the steady-state limit for the non-degenerate double dot coupled
asymmetrically to the leads.

\section{Real-time dynamics}

Having the analytical solution for the electronic occupations and the transport
currents in the double-dot AB interferometer for an initial empty
state, we now examine the real-time dynamics of the electrons for
various values of the on-site energy splitting, the coupling asymmetry as well as the
externally-applied bias.  For simplicity, we exploit the bias configuration, 
$\mu_{L}=eV/2=-\mu_{R}$.

\subsection{Degenerate double dot with asymmetric couplings to the leads ($\delta{E}= 0$ but
$\delta\Gamma \ne 0$)}

When $\delta{E}=0$, one can see from Eq.~(\ref{ud}) that the time
needed to reach the steady-state limit becomes considerably longer as
$\phi$ approaches zero.  This is because the dominant decay factor
given by $\gamma^-_\phi= \frac{1}{2}(\Gamma-\Gamma_\phi)$ in
Eq.~(\ref{ud}) becomes smaller as $\Gamma_\phi =\sqrt{\Gamma^2
\cos^2(\phi/2)+\delta \Gamma^2 \sin^2 (\phi/2)} > 0$ becomes larger,
when $\phi$ approaches zero. However, at $\phi=0$,
$\Gamma-\Gamma_{\phi=0}=0$ and the time to reach the steady-state
limit, given by $(\gamma^-_{\phi=0})^{-1}=
\frac{1}{2}(\Gamma+\Gamma_{\phi=0})^{-1}=\Gamma^{-1}$, becomes much
shorter. This is because the applied magnetic flux breaks the
occupation symmetry associated with the degeneracy, as we have
discussed at the end of Sec.~IV. As a result, the total
occupation is discontinuous across the zero flux point at
steady state. In the time domain, this effect is manifested as the
apparent elongation of the time scale for reaching the steady-state
limit at small but nonzero fluxes.  The nonequilibrium occupation
dynamics of the system with and without a threading magnetic flux
becomes therefore significantly different.

\begin{figure}[h]
\includegraphics[width=8cm, height=3.cm]{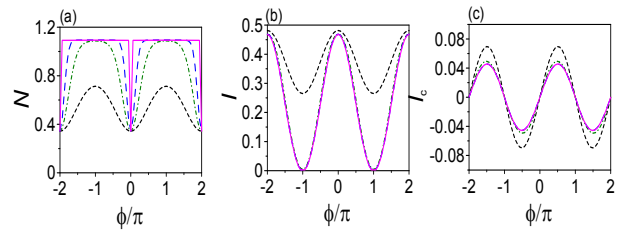}
\caption{Flux dependence of the occupations and currents at several
different times for $\delta{E}=0$.  The quantum dots are initially
empty.  The difference in the occupation numbers is not shown since
it remains zero at degeneracy.  The black short-dashed line is for
$t=2/\Gamma$, the green dash-dotted line is for $t=10/\Gamma$, the
blue long-dashed line is for $t=40/\Gamma$ and $t=\infty$ is the
magenta solid line. The bias is
$eV=3\Gamma$, the asymmetric coupling is $\delta\Gamma=-0.5\Gamma$ and
the temperature is $k_{B}T=\Gamma/20$.  The parameters used here are also used in
other figures unless otherwise stated. \label{fig2} }
\end{figure}

The time-dependent AB oscillations of the total occupation,
the net current and the circulating current are shown 
in Figs.~\ref{fig2}(a), (b) and (c), respectively, for the degenerate double-dot
asymmetrically coupled to the leads. The flux dependencies
$N(\phi,t)=N(-\phi,t)$, $I(\phi,t)=I(-\phi,t)$ as well as
$I_{c}(t)\propto\sin(\phi)$ discussed in Sec.~III are shown there. 
In Fig.~\ref{fig2}(a), we see
that the curve for $t=40/\Gamma$ deviates from the
curve for $t=\infty$ for fluxes near zero.  A discontinuity of the
total occupation across the zero flux point is shown for the curve
at $t=\infty$. This demonstrates the occupation symmetry breaking by
the applied flux at degeneracy as discussed in Sec.~IV. In Sec.~IV,
we have also pointed out that both the net current and the circulating current are continuous
across the zero flux point.  Therefore the long times needed for the occupation 
to reach the steady state near zero flux are not
expected for these currents, as shown in Fig.~\ref{fig2} (b) and (c).

\subsection{Non-degenerate double dot with symmetric coupling to the leads ($\delta{E}\ne0$ but
$\delta\Gamma=0$)}

We proceed to examine the case
with arbitrary on-site energy difference on the two dots  coupled symmetrically to
the leads, i.e. $\delta{E}\ne 0$ but $\delta \Gamma=0$.

\begin{figure}[h]
\includegraphics[width=8cm, height=8.5cm]{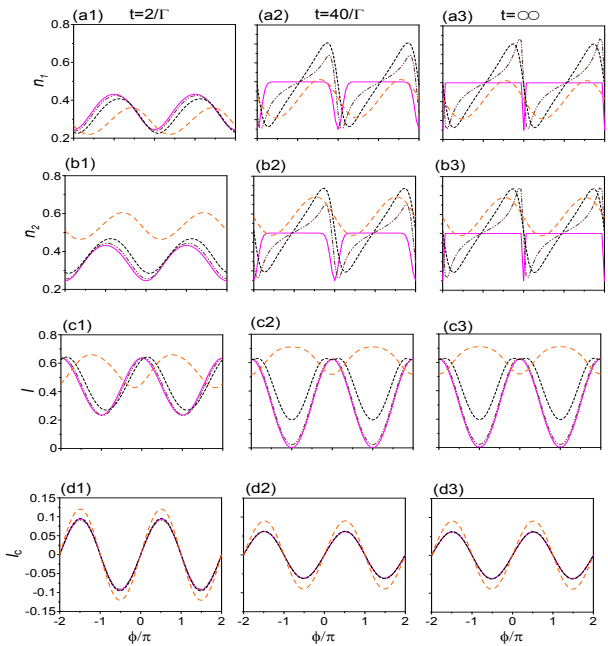}
\caption{Flux dependencies of the occupations and currents at
several different times with various energy splittings $\delta{E}$.
The magenta solid lines are for $\delta{E}=0$, the brown
dash-dot-dot lines are for $\delta{E}=0.15\Gamma$, the black
short-dashed lines are for $\delta{E}=0.5\Gamma$ and the orange
long-dashed lines are for $\delta{E}=2\Gamma$. \label{fig3} }
\end{figure}

The flux dependence profiles of the occupations in each quantum dot,
the net current and the circulating current are plotted in Fig.~\ref{fig3}
for several values of the on-site energy splitting. With a small energy
splitting, for example, $\delta{E}=0.15\Gamma$, we can still see
that the times needed for the occupations to reach the steady state  at fluxes
near zero are much longer than those at other values of the flux
(compared Figs.~\ref{fig3}(a2) and (b2) at $t=40/\Gamma$ with (a3)
and (b3) at $t=\infty$).  However, the occupations at steady
state are continuous across zero flux when the degeneracy
of the double dot system is removed, even only slightly, as shown by the curve for
$\delta{E}=0.15\Gamma$ in Figs.~\ref{fig3}(a3) and (b3).  By further
increasing $\delta E$, the electrons are more likely to
occupy the level with the lower energy as expected, see the curves for
$\delta{E}=2\Gamma$ in Figs.~\ref{fig3}(a2), (a3) and (b2), (b3).
Non-symmetric flux dependencies are also observed for the
occupations at $\delta{E}\ne0$, as discussed in Section III.

On the other hand, upon increasing $\delta{E}$ to a large value (here
$\delta{E} \sim 2\Gamma$), there is a $\pi$ phase jump with respect
to $\delta{E}=0$ in the net current, see the orange long-dashed
lines in Figs.~\ref{fig3}(c1) to (c3). The phase
jump upon changing $\delta{E}$ in the steady-state net current can
be easily found from Eq.~(\ref{sscurrent}), regardless of the value
of $\delta\Gamma$.  More interestingly, the AB oscillation pattern
for different $\delta{E}$ at time $t=2/\Gamma$, see
Fig.~\ref{fig3}(c1), has a different phase shift as compared to those at a later
time, see Figs.~\ref{fig3}(c2) and (c3). This shows the transiently
breaking of the phase rigidity in the transient net current, as we
have discussed in connection with Eq.~(\ref{netcrnt}). In contrast, for the case
of the degenerate double dot, phase rigidity
remains at all times, as shown in Fig.~\ref{fig2}.  The behavior of the circulating
current in response to the change of $\delta{E}$ is rather simple.
Its AB oscillations remain proportional to $\sin\phi$ at all
 times for different $\delta{E}$ with $\delta\Gamma=0$,
see Figs.~\ref{fig3}(d1) to (d3),  as expected from
Eq.~(\ref{circrnt}).

\subsection{Non-degenerate double dot with the asymmetric coupling to the leads ($\delta{E} \ne 0$
and $\delta\Gamma \ne 0$)}

After examining the effects of the asymmetric coupling and the
energy splitting separately on the electronic occupation and transport
dynamics, we next study the effects of the asymmetric coupling
together with a finite energy splitting

\begin{figure}[h]
\includegraphics[width=7.5cm, height=7.0cm]{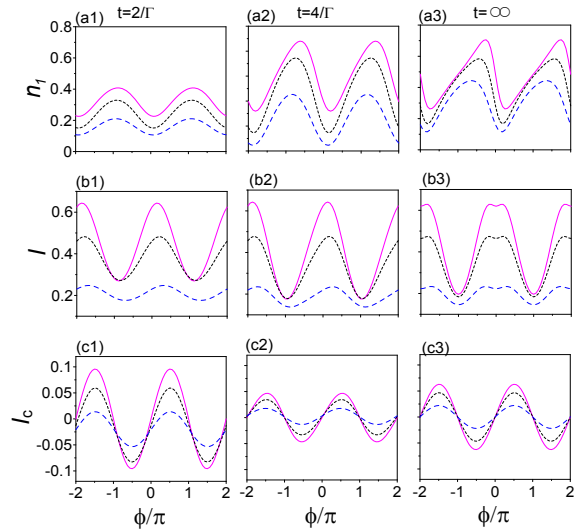}
\caption{The occupation number of the first dot, the net current and
the circulating current as functions of flux in (a1) to (a3), (b1) to (b3) and
(c1) to (c3), respectively, with $\delta{E}=0.5\Gamma$. The magenta solid
lines are for $\delta\Gamma=0$, the black short-dashed lines are for
$\delta\Gamma=-0.5\Gamma$ and the blue long-dashed lines are for
$\delta\Gamma=-0.8\Gamma$.
\label{fig4} }
\end{figure}

Figure \ref{fig4} illustrates the results for the occupations and currents
under various choices of $\delta\Gamma$'s with $\delta{E} =
0.5\Gamma$. The occupation of the second quantum dot is not shown
 for this energy splitting since $n_{2}$ does not differ much from
$n_{1}$.  At time $t=2/\Gamma$, the occupations for different
asymmetries in the coupling have the same AB oscillation phases, see
Fig.~\ref{fig4}(a1).  At long times, the differences in the AB
oscillation phases due to different asymmetric couplings become more
distinct.  This
demonstrates the arbitrary flux dependence of the
AB oscillations of
the total occupation with respect to the change of $\delta\Gamma$.

Figures \ref{fig4}(b1) to (c3) show that the amplitudes of both the
transient net current and circulating current decrease upon
increasing the coupling asymmetry, consistent with what we have
found from Eq.~(\ref{sscurrent}) and Eq.~(\ref{stdy_Ic}). Inspecting
the curves in Fig.~\ref{fig4}(c1), one sees that the flux dependence
of the circulating current has a small deviation from the $\sin\phi$
profile during a short time initially, but then it reaches the
profile proportional to $\sin\phi$.

\begin{figure}[h]
\includegraphics[width=8cm, height=8.5cm]{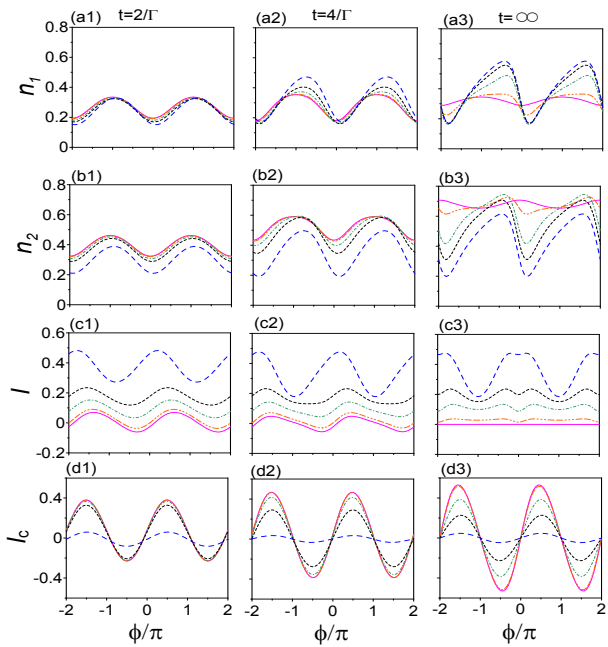}
\caption{The occupation numbers (a1)-(b3), the net
current (c1)-(c3) and the circulating current (d1)-(d3) as functions
of flux. The magenta
solid lines are for $eV=0$, the orange dash-dash-dot-dot-dot lines
are for $eV=0.0125\Gamma$, the green dash-dot lines are for
$eV=0.5\Gamma$, the black short-dashed lines are for $eV=\Gamma$ and
the blue long-dashed lines are for $eV=3\Gamma$. \label{fig5} }
\end{figure}

So far, we have examined the effects of changing the asymmetries of
the system, namely, $\delta{E}$ and $\delta\Gamma$, at a fixed bias.
We now turn to study the effects of varying the bias. The flux
dependence profiles of the electronic occupations for different
choices of the bias are plotted in Figs.~\ref{fig5}(a1) to (b3), for a
given energy splitting and asymmetric coupling, $\delta E=0.5
\Gamma$ and $\delta \Gamma=-0.5 \Gamma$.  When the bias increases,
the difference in the occupations becomes smaller [see the curves
for $eV=\Gamma$ and $3\Gamma$ in Figs.~\ref{fig5}(a2), (a3) and (b2), (b3)].  
The symmetric flux dependence of the net current is transiently
broken for all biases [see Figs.~\ref{fig5}(c1) and (c2)].  At zero
bias, the net current goes to zero at steady state, as expected,
but a finite transient net current is observed [see the curve
for $eV=0$ in Figs.~\ref{fig5}(c1) and (c2)]. At finite biases, when the
net current evolves to its steady-state value, the AB oscillations
develop a sub-oscillatory pattern, proportional to $\sin^{2}(\phi/2)$,
over the main oscillation of $\cos^{2}(\phi/2)$ [see the curve for
$eV=0.0125\Gamma$ in Fig.~\ref{fig5}(c3)]. The anti-symmetric flux
dependence is maintained for the circulating current, only the AB
oscillation amplitudes vary in time for different biases
[see Figs.~\ref{fig5}(d1) and (d3)].

\section{Conclusions}
In this work we have explored the transient quantum
dynamics of a double-quantum-dot AB interferometer using the exact
solution of the  master equation.  We analyzed the effects of
various tunable parameters of the system, namely, the
splitting of the on-site energies on the double dot, the asymmetric coupling to the left
and the right leads and the externally-applied bias, on the time-dependent
electronic occupations and the net current as well as on the circulating
current, during the nonequilibrium transient processes. In the
steady-state limit, we recover the results that have been
extensively investigated in the literature.

With  identical on-site energies on
the double dot, regardless of the coupling asymmetry to the leads, we find
that the total electronic occupation in the double dot and the net
current are always symmetric in the flux, while the occupation
difference between the two dots and the circulating current are
anti-symmetric in it. We also find that the time needed for the
total occupation to reach its steady-state value is much longer
near zero flux, compared with the flux values away from zero. This is because
there exists an occupation symmetry at zero flux, where a
discontinuity cross zero flux in the total occupation is
 found.  By breaking the degeneracy of the double dot, the
phase rigidity in the net current is broken  transiently at an
arbitrary bias. By varying the non-degeneracy of the double dot and the
coupling asymmetry to the leads, the total occupation has
an arbitrary flux dependence at finite biases. The non-degenerate double
dot with an asymmetric coupling to the leads makes the circulating
current to slightly deviate from the anti-symmetric flux dependence 
initially, but it then quickly
approaches the AB oscillations with the fully anti-symmetric flux
dependence. The net current shows a sub-oscillatory pattern
over the main oscillation of $\cos\phi$ at finite bias. It is also
shown that a small bias causes a large circulating current whereas
the net current is negligible. Thus the circulating current may
provide new insights into electron coherence during the
transport.

In short,  the splitting of the on-site energies on the double dot and the 
bias configuration applied to the leads change significantly the flux 
dependencies of the transient electronic occupations as well as the 
transient transport currents.  
We hope that experimentally monitoring the transient behaviors will deepen 
our understanding of the electronic dynamics in quantum-dot AB interferometers.

\begin{acknowledgements}
This work is partially supported by the National Science Council (NSC) of ROC
under Contract No. NSC-99-2112-M-006-008-MY3. We also acknowledge
support from the National Center for Theoretical Science of NSC and the
High Performance Computing Facility in the National Cheng Kung
University. OEW and AA acknowledge support from the Israel Science
Foundation.
\end{acknowledgements}

\end{document}